\documentclass[journal]{IEEEtran}
\usepackage{amsmath}
\usepackage{cite}
\usepackage{amsfonts,amssymb}
\usepackage{graphicx}
\usepackage{subfigure}
\usepackage{multirow}
\usepackage{enumerate}
\usepackage[shortlabels]{enumitem}
\usepackage[T1]{fontenc}
\usepackage{stfloats}
\usepackage{booktabs}
\usepackage[noend]{algorithmic}
\usepackage{tikz}
\usepackage[subnum]{cases}
\usepackage{algorithm}
\usepackage{array}
\usepackage{color}
\usepackage{colortbl}
\usepackage{mathtools}
\usepackage{hyperref}

\usepackage{balance}

\usetikzlibrary{trees}

\ifCLASSINFOpdf
\else
\fi
\algsetup{indent=2em}

\floatname{algorithm}{\bf{Summary of the proposed algorithm}}
\hypersetup{hidelinks}

\begin{document}
%
\title{Blind Identification of SFBC-OFDM Signals Using Subspace Decompositions and Random Matrix Theory}
%
%
%

\author{
        Mingjun Gao, \IEEEmembership{Student Member,~IEEE,}
        Yongzhao Li, \IEEEmembership{Senior Member,~IEEE,}\\
        Octavia A. Dobre, \IEEEmembership{Senior Member,~IEEE,} and
        Naofal Al-Dhahir, \IEEEmembership{Fellow,~IEEE} 
}

\maketitle

\begin{abstract}
Blind signal identification has important applications in both civilian and military communications. Previous investigations on blind identification of space-frequency block codes (SFBCs) only considered identifying Alamouti and spatial multiplexing transmission schemes. In this paper, we propose a novel algorithm to identify SFBCs by analyzing discriminating features for different SFBCs, calculated by separating the signal subspace and noise subspace of the received signals at different adjacent OFDM sub-carriers. Relying on random matrix theory, this algorithm utilizes a serial hypothesis test to determine the decision boundary according to the maximum eigenvalue in the noise subspace. Then, a decision tree of a special distance metric is employed for decision making. The proposed algorithm does not require prior knowledge of the signal parameters such as the number of transmit antennas, channel coefficients, modulation mode and noise power. Simulation results verify the viability of the proposed algorithm for a reduced observation period with an acceptable computational complexity.
\end{abstract}


%
\IEEEpeerreviewmaketitle

\section{Introduction}
%
%
%
%
\IEEEPARstart{D}{ue} to its increasing civilian and military applications, blind identification of communication signal parameters without reference signals has received increased attention recently. Military applications include blind identification of potentially hostile communication sources in radio surveillance, interference identification, electronic warfare and forensics for securing wireless communications\cite{Survey_Signal_Identification,Sec_physical_Layer}. In the context of civilian use, employing blind identification algorithms at the receiver is critical for software defined radios and cognitive radios to improve power and spectral efficiencies \cite{Survey_Signal_Identification}. Recently, numerous algorithms have been developed for the blind identification of multiple-input-multiple-output (MIMO) signal parameters such as the number of transmit antennas \cite{AIC_MDL,WME,HOM_TD_Nt_est} and space-time block codes (STBC) \cite{Likelihood_Based,correlator_function,higher_order_cyclic,blind_recognition_STBC,Hierarchical_STBC, Fourth_order_TC,Second_Order_cyclic,K_S_test,Classify_STBC_Over_FS,STBC_cyclic_2015_ICC,Blind_MIMO_OFDM,Blind_MIMO_OFDM_SM_AL,Identification_SM_AL_OFDM_cyclic,blind_SFBC,My_paper_WCNC,My_paper_Globecom}.

\begin{figure*}
  \centering
  \includegraphics[width=1\textwidth]{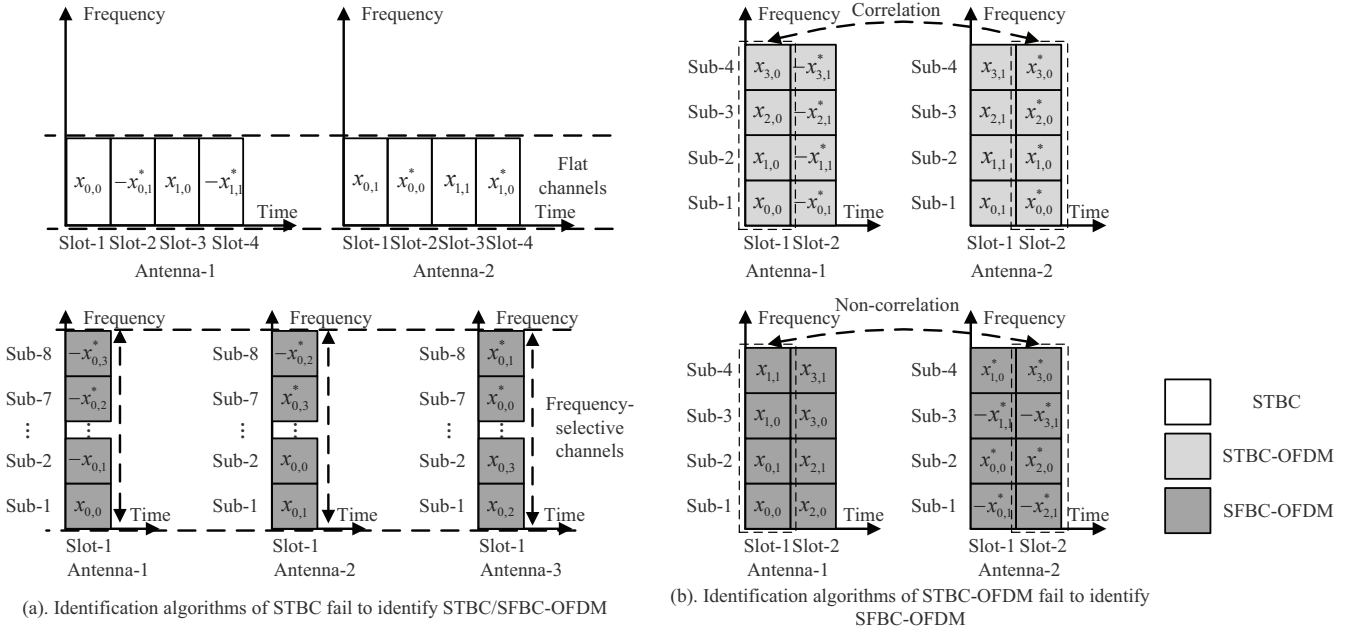}\\
  \caption{Differences among the identification of STBC, STBC-OFDM and SFBC-OFDM.}\label{fig1}
\end{figure*}

Previously reported investigations on the identification of STBC include references \cite{Likelihood_Based,correlator_function,higher_order_cyclic,blind_recognition_STBC,Hierarchical_STBC, Fourth_order_TC,Second_Order_cyclic,K_S_test,Classify_STBC_Over_FS,STBC_cyclic_2015_ICC} for single-carrier systems and references \cite{Blind_MIMO_OFDM,Blind_MIMO_OFDM_SM_AL,Identification_SM_AL_OFDM_cyclic,blind_SFBC,My_paper_WCNC,My_paper_Globecom} for orthogonal frequency division multiplexing (OFDM) systems. Regarding the identification of STBC for single-carrier systems, previous works can be divided into two types of algorithms: likelihood-based\cite{Likelihood_Based} and feature-based \cite{correlator_function,higher_order_cyclic,blind_recognition_STBC,Hierarchical_STBC, Fourth_order_TC,Second_Order_cyclic,K_S_test,Classify_STBC_Over_FS,STBC_cyclic_2015_ICC} algorithms. All of these algorithms are not applicable to OFDM systems over frequency selective fading channels as shown in Fig. \ref{fig1}.a. As for OFDM systems, there are two major spatial transmit diversity approaches. The first is STBC-OFDM which implements the spatial redundancy over adjacent OFDM symbols and has been adopted in indoor WiFi standards \cite{IEEE802_11,STBC_on_80211}. However, under high mobility scenarios, implementing the STBC over adjacent OFDM symbols is ineffective due to the significant channel time variations. Instead, another spatial transmit diversity approach, namely, space-frequency block code (SFBC), is considered where the spatial redundancy is implemented over adjacent OFDM sub-carriers within the same OFDM symbol. Several wireless standards, such as LTE \cite{sesia2009lte} and WiMAX \cite{IEEE802_16,STBC_4_ant}, have adopted SFBC-OFDM. In \cite{Blind_MIMO_OFDM,Blind_MIMO_OFDM_SM_AL,Identification_SM_AL_OFDM_cyclic}, the authors proposed detecting the peak of the cross-correlation function in the time-domain to identify STBC-OFDM signals. However, the time-domain cross-correlation between adjacent OFDM symbols does not exist any longer for SFBC-OFDM signals. Thus, blind identification algorithms of STBC-OFDM cannot be directly applied to SFBC-OFDM signals as shown in Fig. \ref{fig1}.b. The authors of \cite{blind_SFBC,My_paper_Globecom} apply the principle of STBC-OFDM identification to the SFBC scenario. These algorithms detect the peak of the cross-correlation in one OFDM symbol. However, they can only identify a small number of SFBCs due to the identical location of the peak for many SFBCs. To tackle this challenge, our prior work \cite{My_paper_WCNC} used quantified features to make SFBCs distinguishable, nevertheless, it has low performance for low signal-to-noise ratio (SNR) and higher computational complexity.

In order to improve the performance and reduce the complexity, in this paper, we propose an extended SFBC identification algorithm for MIMO-OFDM transmissions over frequency-selective channels. First, we derive a discriminating feature vector for different SFBCs by analyzing the signal subspace and noise subspace of the received signals at adjacent OFDM sub-carriers. Then, the discriminating vector is calculated via a serial binary hypothesis test based on an asymptotically accurate expression from random matrix theory (RMT). Furthermore, we propose a decision tree based scheme which uses a special distance metric to provide a better identification performance with a short observation period in the low SNR range and reduce the computational complexity. The proposed algorithm does not require \textit{a priori} knowledge of signal parameters such as the number of transmit antennas, channel coefficients, modulation mode and noise power. In addition, the proposed algorithm can identify single-antenna (SA) and spatial multiplexing (SM) signals.

The main contributions of this paper are summarized as follows:
\begin{itemize}
\item
The proposed algorithm improves the performance of \cite{My_paper_WCNC} by using an asymptotically accurate expression and a decision tree with a special distance metric.
\item
The proposed algorithm reduces the computational complexity of \cite{My_paper_WCNC} by taking advantage of the tree's decision structure.
\item
The performance of the proposed algorithm is analyzed. An expression for a weak upper bound on the probability of correct identification is derived, and the consistency of the proposed algorithm is proved.
\item
Simulation results are presented to demonstrate the viability of the proposed algorithm with a short observation period in the low SNR range.
\end{itemize}

This paper is organized as follows. In Section II, the signal model is introduced. Section III describes the proposed algorithm, and in Section IV, the simulation setup and results are presented. Finally, conclusions are drawn in Section V.

{\bf Notation:} Following notation is used throughout the paper. The superscripts ${   (\cdot )  ^ * }$, ${   (\cdot)  ^ {T} }$ and ${   (\cdot)   ^ {H} }$ denote complex conjugate, transposition and conjugate transposition, respectively.  $\Pr \left\{ B \right\}$ represents the probability of the event $B$. $\Pr \left\{ B|A \right\}$ represents the conditional probability of the event $B$ under the condition $A$. ${\rm{E}}\left[  \cdot  \right]$ indicates statistical expectation. A complex value can be expressed as $\Re \left(  \cdot  \right) + j \cdot \Im \left(  \cdot  \right)$, where $\Re \left(  \cdot  \right)$ and $\Im \left(  \cdot  \right)$ denote the real and imaginary parts, respectively, and $j^2 = -1$.  $\bf{I}$ denotes the identity matrix. $\mathbb{N}$ and $\mathbb{Z}^+$ denote the set of natural numbers and positive integer, respectively. The notation ${d^{\left( v \right)}}$ represents the symbol $d$ at the $v$-th transmit or receive antenna.

\section{Preliminaries}

\subsection{Conventional Identification of STBC}

The identification of STBC is the process of classifying the SM or STBC signals, which utilizes space-time redundancy to reduce the error rate. From a practical point of view, STBC has three forms, including single-carrier STBC, STBC-OFDM and SFBC-OFDM. For single-carrier systems, the STBC encoder takes the row of an $N_t \times L$ STBC codeword matrix to span $L$ consecutive time slots and maps every column of the matrix into $N_t$ different transmit antennas, where the redundancy is between consecutive time slots. In STBC-OFDM, the STBC codeword is implemented at the same sub-carriers of consecutive OFDM symbols. Different from STBC-OFDM, the SFBC-OFDM encoder takes the row of the codeword matrix to span $L$ consecutive sub-carriers directly, where the redundancy is between consecutive sub-carriers. Unfortunately, the existing algorithms of single-carrier STBC and STBC-OFDM can not be directly employed to identify SFBC-OFDM since the three forms of STBC have different received signal waveforms. Fig. \ref{fig1} (a) shows that the identification algorithms of single-carrier STBC fail to identify STBC/SFBC-OFDM signals since the received signals are degraded by multipath effects. Fig. \ref{fig1} (b) shows that the identification algorithms of STBC-OFDM can not identify SFBC-OFDM either since the signals between consecutive time slots are non-correlated in SFBC-OFDM.

\subsection{System Model}

We consider a MIMO-OFDM wireless communication system with ${N_t}$ transmit antennas and ${M}$-PSK or ${M}$-QAM signal constellations ($M \ge 4$). Here, the transmitted symbols are assumed to be independent and identically distributed (i.i.d.), and the average modulated symbol energy is normalized to one. Subsequently, the modulated data symbol stream is parsed into data blocks of ${N_s}$ symbols, denoted by ${{\bf{x}}_b} = {\left[ {{x_{b,0}}, \cdots ,{x_{b,{N_s} - 1}}} \right]^T}$. The SFBC encoder takes a $N_t \times  L$ SFBC codeword matrix, denoted by ${\bf{C}}\left( {{{\bf{x}}_b}} \right)$, to span $L$ consecutive subcarries in an OFDM symbol. The SFBC codeword matrices for SM $\left( {{N_s} = {N_t},L = 1} \right)$ and Alamouti (AL) $\left( {{N_t} = 2,{N_s} = 2,L = 2} \right)$ are shown in \eqref{eq1} and \eqref{eq2}, respectively, below
\begin{align}
& {{\bf{C}}^{{\rm{SM}}}}\left( {{{\bf{x}}_b}} \right) = {\left[ {{x_{b,0}}, \cdots ,{x_{b,{N_t} - 1}}} \right]^T} \label{eq1} \\
& {{\bf{C}}^{{\rm{AL}}}}\left( {{{\bf{x}}_b}} \right) = \left[ {\begin{array}{*{20}{c}}
{{x_{b,0}}}&{{x_{b,1}}}\\
{ - x_{b,1}^ * }&{x_{b,0}^ * }
\end{array}} \right]. \label{eq2}
\end{align}
For SA, the matrix can be seen as SM with ${N_t}=1$. The matrices for SFBC$^{(1)}$ $\left( {{N_t} = 3,{N_s} = 4,L = 8} \right)$ defined in \cite{STBC_Tarokh},  SFBC$^{(2)}$ $\left( {{N_t} = 2,{N_s} = 3,L = 4} \right)$, and SFBC$^{(3)}$ $\left( {{N_t} = 2,{N_s} = 3,L = 4} \right)$ defined in \cite{Larsson_STBC} are given in the Appendix. The $v$-th row of the codeword matrix is transmitted from the $v$-th antenna. The symbols on each antenna are input to the $N$ consecutive sub-carriers of one OFDM block denoted by
\begin{equation}
{\bf{S}}\left( {{{\bf{x}}_b}, \cdots ,{{\bf{x}}_{b + \frac{N}{L} - 1}}} \right) = \left[ {{\bf{C}}\left( {{{\bf{x}}_b}} \right), \cdots ,{\bf{C}}\left( {{{\bf{x}}_{b + \frac{N}{L} - 1}}} \right)} \right]. \label{eq3}
\end{equation}
Then, an OFDM modulator generates the time-domain block, i.e., OFDM symbol, via $N$-point inverse fast Fourier transform and adds the last $\nu$ samples as a cyclic prefix.

At the receiver side, we assume an advanced receiver composed of ${N_r}\left( {{N_r} > {N_t}} \right)$ antennas and perfect synchronization. Later on in the paper, the effect of imperfect synchronization will be discussed. Here, the synchronization parameters, including the starting time of the OFDM symbol, number of sub-carriers and CP length, are assumed to be estimated successfully and fed to the OFDM demodulator. Several blind synchronization algorithms, even for the relatively low-SNR regime by using the cyclostationarity principles, were described in \cite{Cyclic_of_OFDM,OFDM_parameters,OFDM_parameters2}. The received OFDM symbols are first stripped of the cyclic prefix and then converted into the frequency-domain via an $N$-point fast Fourier transform by the OFDM demodulator. We can construct an ${N_t} \times 1$ transmitted signal vector, transmitting one column of  ${\bf{S}}\left( {{\bf{x}}_b}, \cdots, {{\bf{x}}_{b + N /L - 1}} \right)$, denoted by $\mathbf{s}_k=\left[ s_{k}^{\left( 1 \right)},s_{k}^{\left( 2 \right)},\cdots ,s_{k}^{\left( N_t \right)} \right] ^T$, and a ${N_r} \times 1$ received signal vector $\mathbf{y}_k=\left[ y_{k}^{\left( 1 \right)},y_{k}^{\left( 2 \right)},\cdots ,y_{k}^{\left( N_t \right)} \right] ^T$ at the $k$-th OFDM sub-carrier ($1 \le k \le N$).
The channel is assumed to be a frequency-selective fading channel and the $k$-th subchannel is characterized by an ${N_r} \times {N_t}$ full-rank matrix of fading coefficients denoted by
\begin{equation}
{{\bf{H}}_k} = \left[ {\begin{array}{*{20}{c}}
{h_k^{\left( {1,1} \right)}}& \cdots &{h_k^{\left( {{N_t},1} \right)}}\\
 \vdots & \ddots & \vdots \\
{h_k^{\left( {1,{N_r}} \right)}}& \cdots &{h_k^{\left( {{N_t},{N_r}} \right)}}
\end{array}} \right] \label{eq4}
\end{equation}
where $h^{\left( {v_1},{v_2} \right)}$ represents the channel coefficient between the $v_1$-th transmit antenna and the $v_2$-th receive antenna. Then, the $n$-th ($n \in \mathbb{N}$) received signal at the $k$-th OFDM sub-carrier is described by the following signal model
\begin{equation}
{\bf{y}}_k\left( n \right) = {{\bf{H}}_k}  {\bf{s}}_k\left( n \right) + {\bf{w}}_k\left( n \right) \label{eq5}
\end{equation}
where the ${N_r} \times 1$ vector ${\bf{w}}_k\left( n \right)$ represents the complex additive white Gaussian noise (AWGN) at the $k$-th OFDM sub-carrier with zero mean and covariance matrix $\sigma _w^2{{\bf{I}}_{{N_r}}}$.

\section{Proposed Blind SFBC Identification Algorithm}
In this section, the signals at adjacent OFDM sub-carriers are analyzed firstly. Subsequently, the dimension of the signal subspace at adjacent OFDM sub-carriers is used as the discriminating feature for different SFBCs. Using a sliding window in the frequency domain, a discriminating vector is constructed to identify SFBCs. For the estimation of the dimension, we employ a serial binary hypothesis test with an asymptotically accurate expression based on RMT to detect the maximum eigenvalue in the noise subspace. Finally, a decision tree combined with a special distance metric between the estimated discriminating vector and the theoretical one is proposed to compute the result.

\subsection{Discriminating Feature}
Although OFDM signals propagate through frequency selective fading channels, we can reasonably assume that adjacent subchannels degenerate to a flat fading channel since the severity of the fading at adjacent subchannels is virtually identical. Then, we have
\begin{equation}
\mathbf{H}_{k+1}=\mathbf{H}_k+\varDelta \mathbf{H}\approx \mathbf{H}_k \label{eq6}
\end{equation}
where $\varDelta \mathbf{H}$ is a small difference. Let us define the $n$-th transmitted block at the $k$-th pair of adjacent OFDM sub-carrier, denoted by an ${N_t} \times 2$ matrix $\mathbf{S}_k\left( n \right) =\left[ \mathbf{s}_k\left( n \right), \mathbf{s}_{k+1}\left( n \right) \right] $. The $n$-th received block at the $k$-th pair of adjacent OFDM sub-carrier is expressed as
\begin{equation}
\mathbf{Y}_k\left( n \right) =\mathbf{H}_k \mathbf{S}_k\left( n \right) +\mathbf{W}_k\left( n \right) \label{eq7}
\end{equation}
where the noise block is $\mathbf{W}_k\left( n \right) =\left[ \mathbf{w}_k\left( n \right),\mathbf{w}_{k+1}\left( n \right) \right] $.
Let us define a vector which only contains independent symbols, denoted by $\mathbf{\bar{x}}=\left[ x_1,\cdots ,x_m \right] ^T$, i.e., all the elements in vector $\mathbf{\bar{x}}$ are independent from each other.
Then, $\mathbf{S}_k\left( n \right)$ can be alternatively expressed as follows
\begin{equation}
\mathbf{S}_k\left( n \right) =\left[ \mathbf{A}_1\left(k \right)\mathbf{\tilde{x}}_k\left( n \right),\mathbf{A}_2\left(k \right)\mathbf{\tilde{x}}_{k}\left( n \right) \right] \label{eq8}
\end{equation}
where the matrix $\mathbf{A}$ is a symbol generator matrix and the $2N_t$ vector $\mathbf{\tilde{x}}_k\left( n \right)$ is 
\begin{equation}
\mathbf{\tilde{x}}_k\left( n \right) =\left[ \Re \left( \mathbf{\bar{x}}_{k}^{T}\left( n \right) \right),\Im \left( \mathbf{\bar{x}}_{k}^{T}\left( n \right) \right) \right] ^T. \label{eq9}
\end{equation} 
For example, an AL block is transmitted at the $k$-th sub-carrier and its neighbor. Hence, the vector of independent symbols at the $k$-th and $(k+1)$-th OFDM sub-carrier pairs is ${\mathbf{\bar{x}}_{k}} = {\left[ {{x_1},{x_2}} \right]^T}$ and the symbol matrices ${{{\bf{A}}_1}}\left(k \right)$ and ${{{\bf{A}}_2}}\left(k \right)$ are
\begin{equation}
{{\bf{A}}_1}\left(k \right) = \left[ {\begin{array}{*{20}{c}}
1&0&j&0\\
0&{ - 1}&0&j
\end{array}} \right] \  {{\bf{A}}_2}\left(k \right) = \left[ {\begin{array}{*{20}{c}}
0&1&0&j\\
1&0&{ - j}&0
\end{array}} \right].  \label{eq10}
\end{equation}
Another example is two AL blocks transmitted at adjacent OFDM sub-carriers, i.e., the second column of the former block transmitting at $k$-th OFDM sub-carrier and the first column of the latter block transmitting at $(k+1)$-th OFDM sub-carrier. In this case, the vector of independent symbols at the $k$-th and $(k+1)$-th OFDM sub-carrier pairs is $\mathbf{\bar{x}}_k=\left[ x_1,x_2,x_3,x_4 \right] ^T$, respectively. The symbol matrices  are
\begin{align}
\mathbf{A}_1\left( k \right) &=\left[ \begin{matrix}
	0&		1&		0&		0&		0&		j&		0&		0\\
	1&		0&		0&		0&		-j&		0&		0&		0\\
\end{matrix} \right]  \notag \\
\mathbf{A}_2\left( k \right) &=\left[ \begin{matrix}
	0&		0&		1&		0&		0&		0&		j&		0\\
	0&		0&		0&		-1&		0&		0&		0&		j\\
\end{matrix} \right].  
\label{eq11}
\end{align}
By stacking the real and imaginary parts of the signals in \eqref{eq7}, we obtain 
\begin{equation}
\left[ \begin{array}{c}
	\Re \left( \mathbf{Y}_k\left( n \right) \right)\\
	\Im \left( \mathbf{Y}_k\left( n \right) \right)\\
\end{array} \right] ={\mathbf{\bar{H}}}_k\left[ \begin{array}{c}
	\Re \left( \mathbf{S}_k\left( n \right) \right)\\
	\Im \left( \mathbf{S}_k\left( n \right) \right)\\
\end{array} \right] +\left[ \begin{array}{c}
	\Re \left( \mathbf{W}_k\left( n \right) \right)\\
	\Im \left( \mathbf{W}_k\left( n \right) \right)\\
\end{array} \right] \label{eq12}
\end{equation}
where the $2{N_r} \times 2{N_t}$ matrix ${{\bf{\bar H}}_k}$ is given by
\begin{equation}
{{\bf{\bar H}}_k} = \left[ {\begin{array}{*{20}{c}}
{\Re \left( {{\bf{H}}_k} \right)}&{ - \Im \left( {{\bf{H}}_k} \right)}\\
{\Im \left( {{\bf{H}}_k} \right)}&{\Re \left( {{\bf{H}}_k} \right)}
\end{array}} \right]. \label{eq13}
\end{equation}
Then, denote the transmitted block in \eqref{eq12} as a column vector ${\bf{\tilde{s}}}_k\left( n \right)$ of size $4{N_t}$, which is defined as
\begin{equation}
\mathbf{\tilde{s}}_k\left( n \right) =\text{vec}\left\{ \begin{array}{c}
	\Re \left( \mathbf{S}_k\left( n \right) \right)\\
	\Im \left( \mathbf{S}_k\left( n \right) \right)\\
\end{array} \right\}. \label{eq14}
\end{equation}
Denote the received block and noise block as column vectors $\mathbf{\tilde{y}}_k\left( n \right)$, $\mathbf{\tilde{w}}_k\left( n \right)$ of size $4{N_r}$, which are respectively defined as
\begin{subequations} \label{eq15}
\begin{align}
&\mathbf{\tilde{y}}_k\left( n \right) =\text{vec}\left\{ \begin{array}{c}
	\Re \left( \mathbf{Y}_k\left( n \right) \right)\\
	\Im \left( \mathbf{Y}_k\left( n \right) \right)\\
\end{array} \right\}  \label{eq15a}\\
&\mathbf{\tilde{w}}_k\left( n \right) =\text{vec}\left\{ \begin{array}{c}
	\Re \left( \mathbf{W}_k\left( n \right) \right)\\
	\Im \left( \mathbf{W}_k\left( n \right) \right)\\
\end{array} \right\} \label{eq15b}
\end{align}
\end{subequations}
where ${\rm{vec}}\left\{  \cdot  \right\}$ represents vectorization. Under these notations, Equation \eqref{eq12} is finally expressed as
\begin{equation}
\mathbf{\tilde{y}}_k\left( n \right) =\left( \mathbf{I}_2\otimes \mathbf{\bar{H}}_k \right) \mathbf{\tilde{s}}_k\left( n \right) +\mathbf{\tilde{w}}_k\left( n \right)  \label{eq16}
\end{equation}
where $ \otimes $ denotes the Kronecker product.
The covariance matrix ${\mathbf{\Sigma}}_k$ of $\mathbf{\tilde{y}}_k\left( n \right)$ is
\begin{align}
{\mathbf{\Sigma}}_k = &\; \text{E}\left[ \mathbf{\tilde{y}}_k\left( n \right) \mathbf{\tilde{y}}_{k}^{T}\left( n \right) \right] \notag \\
	        = &\left( \mathbf{I}_2\otimes \mathbf{\bar{H}}_k \right) \text{E}\left[ \mathbf{\tilde{s}}_k\left( n \right) \mathbf{\tilde{s}}_{k}^{T}\left( n \right) \right] \left( \mathbf{I}_2\otimes \mathbf{\bar{H}}_k^T \right) \notag \\
	      &+\text{E}\left[ \mathbf{\tilde{w}}_k\left( n \right) \mathbf{\tilde{w}}_{k}^{T}\left( n \right) \right]. \label{eq17}
\end{align}
Next, assume that $m_k$ is the number of independent symbols of $\mathbf{\tilde{x}}_k\left( n \right)$. Since the transmitted symbol energy is normalized, we obtain
\begin{equation}
\text{E}\left[ \mathbf{\tilde{x}}_k\left( n \right) \mathbf{\tilde{x}}_{k}^{T}\left( n \right) \right] =\frac{1}{2}\mathbf{I}_{2m_k}. \label{eq19}
\end{equation}
Additionally, the covariance of the noise is
\begin{equation}
\text{E}\left[ \mathbf{\tilde{w}}_k\left( n \right) \mathbf{\tilde{w}}_{k}^{T}\left( n \right) \right] =\frac{\sigma _{w}^{2}}{2}\mathbf{I}_{4N_r}. \label{eq20}
\end{equation}
As a result, from \eqref{eq17}, \eqref{eq19} and \eqref{eq20}, $\mathbf{\Sigma }_k$ is given as
\begin{equation}
\mathbf{\Sigma }_k=\frac{1}{2}\left( \mathbf{I}_2\otimes \mathbf{\bar{H}}_k \right) \mathbf{M}_k\mathbf{M}_{k}^{T}\left( \mathbf{I}_2\otimes \mathbf{\bar{H}}_{k}^{T} \right) +\frac{\sigma _{w}^{2}}{2}\mathbf{I}_{4N_r} \label{eq21}
\end{equation}
where the matrix $\mathbf{M}_k$ is
\begin{equation}
{\mathbf{M}_k} = \left[ {\begin{array}{*{20}{c}}
{\Re \left( {{{\bf{A}}_1}\left(k \right)} \right)}\\
{\Im \left( {{{\bf{A}}_1}\left(k \right)} \right)}\\
{\Re \left( {{{\bf{A}}_2}\left(k \right)} \right)}\\
{\Im \left( {{{\bf{A}}_2}\left(k \right)} \right)}
\end{array}} \right]. \label{eq22}
\end{equation}
It is easy to verify that the rank of $\left( {{{\bf{I}}_2} \otimes {\bf{\bar H}}} \right)$ is full. We denote the eigenvalues of the covariance matrix $\mathbf{\Sigma }_k$ as $\lambda _1\left( k \right) \ge \cdots \ge \lambda _{4N_r}\left( k \right) $.

\emph{Proposition:} The smallest $4{N_r} - 2m_k$ ordered eigenvalues of $\mathbf{\Sigma }_k$ are all equal to ${\sigma _{w}^{2}}/{2}$, i.e.,
\begin{equation}
{\lambda _{2m_k + 1}}\left( k \right) =  \cdots  = {\lambda _{4{N_r}}}\left( k \right) = {\sigma _{w}^{2}}/{2}. \label{eq23}
\end{equation}

\emph{Proof:}  The rank of $\mathbf{M}_k\mathbf{M}_{k}^{T}$ can be easily shown to be $ 2m_k$, which makes the rank of the first term at the right hand side of \eqref{eq21} equal to $ 2m_k$. The smallest $4{N_r} - 2m_k$ ordered eigenvalues of $\left( \mathbf{I}_2\otimes \mathbf{\bar{H}}_k \right) \mathbf{M}_k\mathbf{M}_{k}^{T}\left( \mathbf{I}_2\otimes \mathbf{\bar{H}}_{k}^{T} \right)$ are equal to zero. Therefore, all of the smallest $4{N_r} -  2m_k$ ordered eigenvalues of $\mathbf{\Sigma }_k$ are equal to ${{\sigma _w^2}}/{2}$.

Actually, the number of independent symbols at the $k$-th OFDM sub-carrier and its neighbor, $2m_k$, is different for different SFBCs at different adjacent OFDM sub-carrier pairs. This number can be seen as the dimension of the signal subspace of the received signals at adjacent sub-carriers after separating the signal and noise subspace. By sliding a frequency-domain window, we can estimate the number of independent symbols for different adjacent OFDM sub-carrier pairs and then construct a discriminating feature vector as follows. 

\begin{figure}
  \centering
  \subfigure[SM-SFBC]{
  \includegraphics[width=0.38\textwidth]{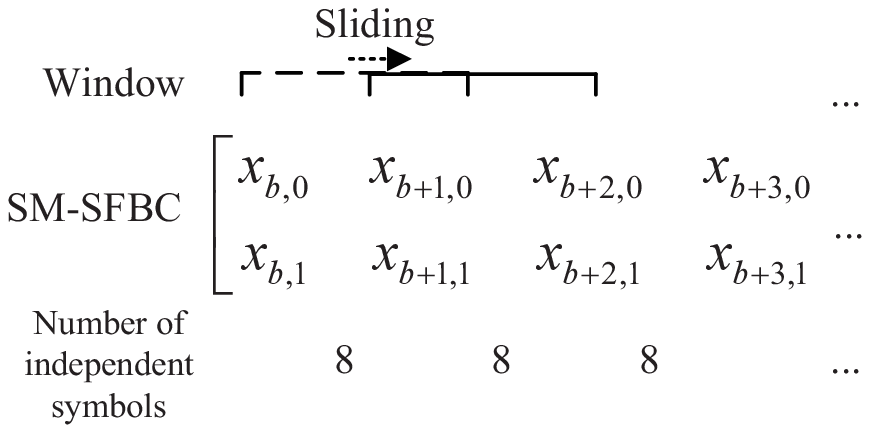}
  }
  \subfigure[AL-SFBC]{
  \includegraphics[width=0.38\textwidth]{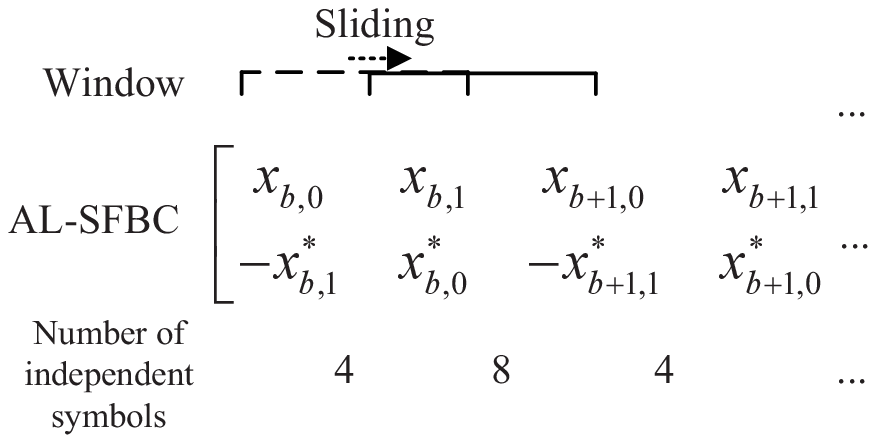}
  }
  \caption{Discriminating feature for SM-SFBC and AL-SFBC.}\label{fig2}
\end{figure}

\subsubsection{\rm SM-SFBC}
Without loss of generality, the case of 2 transmit antennas, ${\rm{SM}}^{\left(2\right)}$, is analyzed first, and the feature vectors of $\rm {SA}$ and ${\rm {SM}}^{\left(3\right)}$ are given afterwards. As shown in Fig. \ref{fig2}.a., the vectors of independent symbols for the first and second OFDM sub-carrier pairs are $\mathbf{\bar{x}}_1=\left[ x_{b,0},x_{b,1} \right] ^T$ and $\mathbf{\bar{x}}_{2}=\left[ x_{b+1,0},x_{b+1,1} \right] ^T$, respectively. Hence, the number at the first pair of adjacent OFDM sub-carriers,  $2m_1$, is equal to 8. By moving the window, the vectors of independent symbols for the second and third OFDM sub-carrier pairs are  $\mathbf{\bar{x}}_{2}=\left[ x_{b+1,0},x_{b+1,1} \right] ^T$ and $\mathbf{\bar{x}}_{3}=\left[ x_{b+2,0},x_{b+2,1} \right] ^T$, respectively. The number $2m_2$ is also equal to 8. Based on sliding the window, we construct a feature vector, denoted by ${\bf{q}}$, whose elements are the numbers $2m_k$. For $\rm{SM}^{(2)}$, the vector is ${{\bf{q}}_{{\rm{S}}{{\rm{M}}^{\left( {\rm{2}} \right)}}}} = \left[ {8,8,8,8,8,8,8,8,\cdots} \right]$. In addition, ${{\bf{q}}_{{\rm{SA}}}} = \left[ {4,4,4,4,4,4,4,4,\cdots} \right]$ is for $\rm {SA}$, and ${{\bf{q}}_{{\rm{S}}{{\rm{M}}^{\left( {\rm{3}} \right)}}}} = \left[ {12,12,12,12,12,12,12,12,\cdots} \right]$ is for ${\rm {SM}}^{\left(3\right)}$, respectively.

\subsubsection{\rm AL-SFBC}
As shown in Fig. \ref{fig2}.b., the vectors of independent symbols for the first and second OFDM sub-carrier pairs are same as mentioned earlier, $\mathbf{\bar{x}}_1=\mathbf{\bar{x}}_{2}=\left[ x_{b,0},x_{b,1} \right] ^T$. Hence, the number, $2m_1$, is equal to 4. After sliding the window to the next OFDM sub-carrier, the vectors of independent symbols for the second and third OFDM sub-carrier pairs are  $\mathbf{\bar{x}}_{2}=\left[ x_{b,0},x_{b,1} \right] ^T$ and $\mathbf{\bar{x}}_{3}=\left[ x_{b+1,0},x_{b+1,1} \right] ^T$, respectively. The number $2m_2$ changes to 8. Consequently, the feature vector is ${{\bf{q}}_{{\rm{AL}}}} = [4,8,4,8,4,8,4,8,\cdots]$.
\subsubsection{\rm Other SFBCs}
Analogously, the feature vector of SFBC$^{(1)}$ is ${{\bf{q}}_{{\rm{SFBC}}^{\rm{(1)}}}} = \left[ {8,8,8,8,8,8,8,12,\cdots} \right]$, that of SFBC$^{(2)}$ is ${{\bf{q}}_{{\rm{SFBC}}^{\rm{(2)}}}} = \left[ {6,6,6,8,6,6,6,8,\cdots} \right]$, and  ${{\bf{q}}_{{\rm{SFBC}}^{\rm{(3)}}}} = \left[ {6,6,6,10,6,6,6,10,\cdots} \right]$ is for SFBC$^{(3)}$, respectively.

\subsection{Classification of the Feature Vectors}
First, we describe the method used to compute the $k$-th element of the estimated feature vector ${\bf{\hat q}}\left(k \right)$. According to \eqref{eq17}, the estimated covariance matrix of the received vectorized signals is given by
\begin{equation}
{{\bf{R}}_{k}} = \frac{1}{{{N_b}}}\sum\limits_{n = 1}^{{N_b}} {{\bf{\tilde y}}_k\left( n \right)} {\bf{\tilde y}}_k{\left( n \right)^T} \label{eq24}
\end{equation}
where ${N_b}$ is the number of OFDM symbols. The eigenvalues of ${{\bf{R}}_{k}}$ are denoted by $l_1{(k)} \ge  \cdots  \ge l_{4{N_r}}{(k)}$, which can be divided into the signal subspace ${L_s} = \{ {l_1{(k)}, \cdots ,l_{2m_k}{(k)}} \}$ and noise subspace ${L_w} = \{ {l_{2m_k + 1}{(k)}, \cdots ,l_{4{N_r}}{(k)}} \}$. From {\emph{Lemma 1}} in \cite{the_num_det_RMT}, when $4{N_r},{N_b} \to \infty ,{{4{N_r}} \mathord{\left/
 {\vphantom {{4{N_r}} {N_b}}} \right.
 \kern-\nulldelimiterspace} {N_b}} \to const > 0$, the eigenvalue $l_{2m_k + 1}{(k)}$ has asymptotically the same Tracy-Widom distribution as the largest eigenvalue of a pure noise Wishart matrix. The noise power ${{\sigma _w^2} \mathord{\left/
 {\vphantom {{\sigma _w^2} 2}} \right.
 \kern-\nulldelimiterspace} 2}$ can be estimated by the average trace of ${L_w}$ as $\frac{1}{4N_r-2m_k}\sum{_{i=2m_k+1}^{4N_r}l_{i}{\left( k \right)}}$. Hence, the test statistic of the $k$-th pair of adjacent OFDM sub-carrier is constructed as
 \begin{equation}
 U_{2m_k+1}{\left( k \right)}=\frac{l_{2m_k+1}{\left( k \right)}}{\frac{1}{4N_r-2m_k}\sum{_{i=2m_k+1}^{4N_r}l_{i}{\left( k \right)}}}. \label{eq25}
 \end{equation}
Consequently, the distribution function of $U_{2m_k + 1}{\left( {k} \right)}$ follows an asymptotically accurate expression as\cite{TW_accrucy}
\begin{align}
& \Pr \left\{ \frac{U_{2m_k + 1}{(k)} - {{\mu} _{4{N_r} - 2m_k,{N_b}}}} {\xi_{4N_r-2m_k, N_b}} \le z  \right\}  \approx  \nonumber \\
& F_{TW1}(z)-\frac{1}{(4N_r-2m_k)N_b} \left( \frac{\mu_{4N_r-2m_k, N_b}}{\xi_{4N_r-2m_k, N_b}} \right)^2 F_{TW1}^{''}(z)\label{eq26}
\end{align}
where ${F_{TW1}}\left(  \cdot  \right)$ and ${F_{TW1}^{''}}\left(  \cdot  \right)$ are the cumulative distribution functions of the Tracy-Widom distribution for the real value noise and its second-order derivative, respectively. The centering and scaling parameters, ${\mu _{u,p}}$ and ${\xi _{u,p}}$, respectively, are given as
\begin{equation}
\left\{ \begin{array}{l}
{\mu _{u,p}} = {\left( {\sqrt {u - 0.5}  + \sqrt {p - 0.5} } \right)^2}\\
{\xi _{u,p}} = \sqrt {{\mu _{u,p}}} {\left( {{1 \mathord{\left/
 {\vphantom {1 {\sqrt {u - 0.5{\rm{ }}} }}} \right.
 \kern-\nulldelimiterspace} {\sqrt {u - 0.5{\rm{ }}} }} + {1 \mathord{\left/
 {\vphantom {1 {\sqrt {p - 0.5} }}} \right.
 \kern-\nulldelimiterspace} {\sqrt {p - 0.5} }}} \right)^{{1 \mathord{\left/
 {\vphantom {1 3}} \right.
 \kern-\nulldelimiterspace} 3}}}
\end{array} \right. \label{eq27}
\end{equation}
where $u$ and $p$ are two parameters of the Wishart distribution. Specifically, $u$ and $p$ are the number of row and column of a random matrix, denoted by $\bf{Y}$, if the Wishart matrix is ${\bf{W}} = {\bf{Y}\bf{Y}}^T$. Then, the number $2m_k$ can be determined by a serial binary hypothesis test. Its decision criterion follows
\begin{equation}
\left\{ {\begin{array}{*{20}{c}}
{{U_q{\left( {k} \right)}} > {\gamma _q},\quad \quad {\rm{under}}\;{{\cal H}_1}}\\
{{U_q{\left( {k} \right)}} \le {\gamma _q},\quad \quad {\rm{under}}\;{{\cal H}_0}}
\end{array}} \right. \label{eq28}
\end{equation}
where $U_q{\left( {k} \right)}$ is the test statistic and ${\gamma _q}$ is the threshold with $q = 1,2, \cdots ,4{N_r}$. The hypothesis ${{\cal H}_1}$ holds when the eigenvalue $l_q{(k)}$ corresponding to $U_q{\left( {k} \right)}$ is a signal eigenvalue (${l_q} \in {L_s}$), while the hypothesis ${{\cal H}_0}$ holds when the eigenvalue $l_q{(k)}$ corresponding to $U_q{\left( {k} \right)}$ is a noise eigenvalue (${l_q} \in {L_w}$). The threshold ${\gamma _q}$ is
\begin{equation}
{\gamma _q} = F_{TW}^{ - 1}\left( {1 - {{\Pr }_f}} \right){\xi _{4{N_r} - q + 1,{N_b}}} + {\mu _{4{N_r} - q + 1,{N_b}}} \label{eq29}
\end{equation}
where $F_{TW}^{ - 1}\left(  \cdot  \right)$ is the inverse function of the right hand side (RHS) of Equation \eqref{eq26} and ${\Pr _f}$ is the false alarm probability. The steps of the test are that we let $q=1,2,\cdots$ and compare $U_q{\left( {k} \right)}$ with ${\gamma _q}$ until the first time that ${{U_q{\left( {k} \right)}} \le {\gamma _q}}$. Then, the $k$-th element of ${\bf{\hat q}}$ is
\begin{equation}
\mathbf{\hat{q}}\left( k \right) =q-1. \label{eq30}
\end{equation}

\begin{figure}
\centering
\begin{tikzpicture}
	\node {SFBC}[edge from parent fork down]
		
		child[xshift=-1.5cm]{
			child[xshift=0.15cm] {node{${\rm{SA}}$}
                             edge from parent
                             node[right,xshift=0.7cm,yshift=0.2cm] {${\bf q}(2i)$}
                             node[right,yshift=-0.25cm] {$=4$}
                              }
			child[xshift=-0.15cm] {node{$\rm{AL}$}
                        edge from parent
                        node[right,yshift=-0.25cm] {$=8$}
                              }
                        edge from parent
                        node[right,yshift=-0.25cm] {$=4$}
			}
		child[xshift=-0.5cm]{
			child[xshift=0.11cm] {node{${\rm SFBC}^{\rm (2) }$}
                             edge from parent
                             node[right,xshift=0.7cm,yshift=0.2cm] {${\bf q}(4i)$}
                             node[right,yshift=-0.25cm] {$=8$}
                              }
			child[xshift=-0.11cm] {node{${\rm SFBC}^{\rm (3) }$}
                        edge from parent
                        node[right,yshift=-0.25cm] {$=10$}
                              }
                        edge from parent
                        node[right,yshift=-0.25cm] {$=6$}
			}
		child[xshift=0.5cm]{
			child[xshift=0.15cm] {node{${\rm{SM}}^{(2)}$}
                             edge from parent
                             node[right,xshift=0.7cm,yshift=0.2cm] {${\bf q}(8i)$}
                             node[right,yshift=-0.25cm] {$=8$}
                              }
			child[xshift=-0.15cm] {node{${\rm SFBC}^{\rm (1) }$}
                        edge from parent
                        node[right,yshift=-0.25cm] {$=12$}
                              }
                        edge from parent
                        node[right,yshift=-0.25cm] {$=8$}
			}	
		child { node[xshift=1.0cm] {${\rm{SM}}^{(3)}$}
                         edge from parent
                         node[right,xshift=-3.1cm,yshift=0.25cm] {${\bf q}(2i-1)$}
                         node[right,yshift=-0.25cm] {$=12$}
                         }

;	
\end{tikzpicture}
\caption{Decision tree for the identification of SFBC.}
\label{Des_tree}

\end{figure}
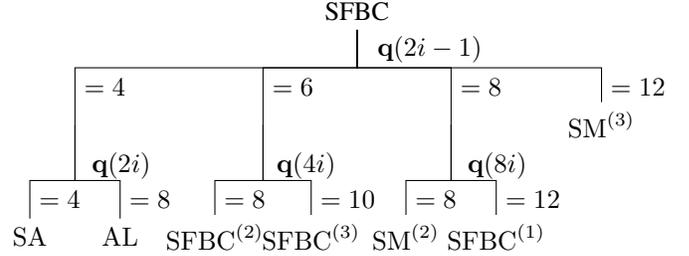

Subsequently, a decision tree classification is proposed to identify different SFBCs, as shown in Fig. \ref{Des_tree}. At the top-level node, we calculate ${\bf{\hat q}}\left(k\right)$ at odd-indexed adjacent sub-carrier pairs, where $k=1,3,\cdots,2i-1,\cdots,N-1$ ($i \in \mathbb{Z}^+  $), and compare the distance, denoted by ${d_c}$, between ${\bf{\hat q}}$ with odd-indexed elements and the theoretical values.
The identified SFBC or subsets, denoted by $\hat C$, is the one which minimizes the distance ${d_c}$ from the set of $ {\rm{SFC}}= \{{\rm SFC}_1, {\rm SFC}_2, {\rm SFC}_3, {\rm SM}^{(3)}  \} $, given as
\begin{equation}
\hat C = \arg \mathop {\min }\limits_{C \in \left\{ {\rm{SFC}} \right\}} {d_c} \label{eq31}
\end{equation}
where the top node yields 4-leaf branches. In this case, subsets ${\rm SFC}_1$, ${\rm SFC}_2$, ${\rm SFC}_3$ and ${\rm SM}^{(3)}$, are given by
\begin{subequations} \label{eq32}
\begin{align}
&{\rm SFC}_1 = \{ \rm{SA},\rm{AL}\} \label{eq32a} \\
&{\rm SFC}_2 = \{ {\rm{SFBC}}^{(2)}, {\rm{SFBC}}^{(3)}\} \label{eq32b}\\
&{\rm SFC}_3 = \{ {\rm SM}^{(3)}, {\rm{SFBC}}^{(1)} \}. \label{eq32c}
\end{align}
\end{subequations}
If the minimum distance is the same for two codes, the code with the smallest ${\bf{q}}\left(k\right)$ is selected. At second-level nodes, subsets ${\rm SFC}_1$, ${\rm SFC}_2$ and ${\rm SFC}_3$ can be divided into corresponding SFBC codes according to  ${\bf{\hat q}}\left(k\right)$ at different sub-carriers. Specifically,  $k=2,4,\cdots,2i,\cdots,N-2$ for the subset  ${\rm SFC}_1$,  $k=4,8,\cdots,4i,\cdots,N-4$ for the subset  ${\rm SFC}_2$ and $k=8,16,\cdots,8i,\cdots,N-8$ for the subset  ${\rm SFC}_3$. Finally, Equation \eqref{eq31} is used to determine the result.

\begin{figure}
  \centering
  \includegraphics[width=0.5\textwidth]{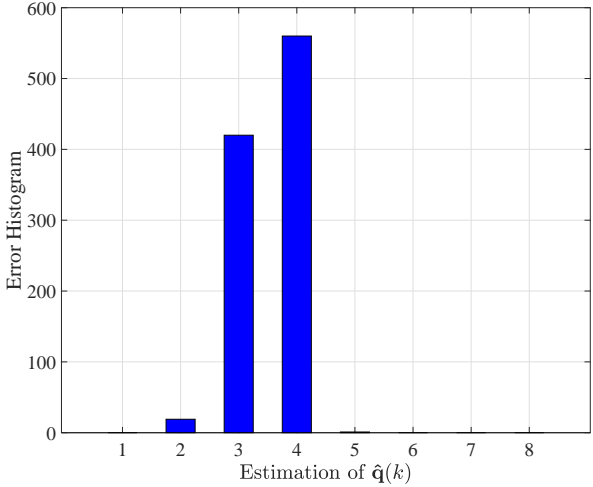}\\
  \caption{Error histogram of the estimation of ${\bf{\hat q}}\left(k\right)$, where correct number is 4 and $N_r=8$, $N_b=400$,  SNR=-1 dB, ${\Pr _f}={10^{ - 4}}$. The simulation was run for 1000 trials. }\label{fig3}
\end{figure}

To improve the performance, we propose to compute a special distance metric ${d_c}$ between ${\bf{\hat q}}$ and the theoretical one for possible sets and codes. The proposed distance ${d_c}$ is
\begin{equation}
d_c=\left| \sum_k{} \varepsilon \left( \mathbf{\hat{q}}\left( k \right) -\mathbf{q}\left( k \right) \right) -\lceil N{\rm{Pr}}_f \rceil \right| \label{eq33}
\end{equation}
where $\left|  \cdot  \right|$ represents the absolute value sign, $\varepsilon \left(  \cdot  \right)$ represents the unit step function, i.e., $\varepsilon \left( t \right) = \left\{ {\begin{array}{*{20}{c}}
{1,t > 0}\\
{0,t \le 0}
\end{array}} \right.$ and $\left\lceil  \cdot  \right\rceil $ indicates the ceiling function. The RHS of the distance formula is explained as follows:
\begin{itemize}[leftmargin=*]
\item \emph{\textbf{The $\sum\limits$ terms}}:
The probability of underestimation when employing the serial hypothesis testing based on RMT is much larger than overestimation probability in the low SNR range, since signal eigenvalues are dominated by the noise. Fig. \ref{fig3} shows that the correct ${\bf{\hat q}}\left(k\right)$ is equal to 4 while the receiver determines a value of 3 far more than the value 5 at SNR = -1 dB. The performance gets worse if we employ the Euclidean distance\cite{My_paper_WCNC} to compare the estimated vector ${\bf{\hat q}}$ with the theoretical one. Therefore, we use the unit step function. Once the estimated value is less than the theoretical one, the term should be set to zero.
\item \emph{\textbf{The last term}}:
Overestimation still occurs due to the setting of ${\Pr _f}$. However, ignoring overestimation will cause non-consensus estimation when processing more sub-carriers in the high SNR range. The conclusion of \cite{TW_accrucy} proves that the expression \eqref{eq26} make ${\Pr _f}$ approximately accurate to describe the probability of overestimation of the test statistic $U_{2m_k + 1}$ for small and even moderate values of $N_r$ and $N_b$, denoted by
\begin{align}
\text{Pr}_o  \approx & \  \text{Pr}_f \nonumber \\ 
\approx & \  1-F_{TW1}\left( \frac{\gamma _q-\mu _{4N_r-2m_k,N_b}}{\xi _{4N_r-2m_k,N_b}} \right)  \nonumber \\
&+\frac{1}{\left( 4N_r-2m_k \right) N_b}\left( \frac{\mu _{4N_r-2m_k,N_b}}{\xi _{4N_r-2m_k,N_b}} \right) ^2 \cdot \nonumber \\
&\ F_{TW1}^{''}\left( \frac{\gamma _q-\mu _{4N_r-2m_k,N_b}}{\xi _{4N_r-2m_k,N_b}} \right). \label{eq34}
\end{align}
Therefore, an error correction factor of ${N{\Pr _f}}$, which represents the times of overestimation during $N$ steps, should be subtracted in case of non-consensus estimations. 
\end{itemize}

The proposed algorithm is summarized as follows
 \begin{algorithm}
 \renewcommand{\thealgorithm}{}
 \caption{}
 \begin{algorithmic}[1]
 \renewcommand{\algorithmicrequire}{\textbf{Input: }}
 \renewcommand{\algorithmicensure}{\textbf{Output:}}
 \REQUIRE The observed symbols sequence $\bf{y}$.
 \ENSURE  7 types of SFBC code $\hat{C}$.
 \STATE set $k=1$
 \REPEAT 
 \STATE{\label{item1}Vectorize the received block of adjacent sub-carriers and get $\mathbf{\tilde{y}}_k$ using \eqref{eq7}, \eqref{eq12} and \eqref{eq15a}} 
 \STATE{Compute the covariance matrix ${{\bf{R}}_{k}}$ using \eqref{eq24}}
 \STATE{Eigenvalue decomposition of ${{\bf{R}}_{k}}$}
 \STATE{Construct the test statistic $U_q{\left( {k} \right)}$ using \eqref{eq25}}
 \STATE{Compute the threshold $\gamma _q$ using \eqref{eq29}}
 \STATE{\label{item2}Estimate ${\bf{\hat q}}\left(k\right)$ using the serial hypothesis test by the decision criterion \eqref{eq28}}
 \STATE{$k=k+2$}
 \UNTIL{$k=N-1$}
 \STATE{Get subsets in \eqref{eq32} \OR continue to the step \ref{item3} with $\hat C = {\rm{SM}}^{(3)}$ using \eqref{eq33} and \eqref{eq31}}
 \STATE According to the previous result, set $k=2$ \OR 4 \OR 8
  \REPEAT 
 \STATE{Step \ref{item1}- Step \ref{item2}} 
 \STATE{$k=k+2$ \OR $k+4$ \OR $k+8$}
 \UNTIL{$k=N-2$ \OR $N-4$ \OR $N-8$}
 \STATE{Obtain $\hat C$ using \eqref{eq33} and \eqref{eq31}}
 \RETURN {\label{item3} $\hat C$}
 \end{algorithmic}
 \end{algorithm}

\subsection{Performance and Consistency of our Algorithm}
The accurate probability of correct identification is difficult to derive due to the heuristic $\sum\limits$ terms of the distance formula in \eqref{eq33}. However, a weak upper bound of the probability on correct identification can be calculated by the probability of overestimation.
The last term of \eqref{eq33} can tolerate $\lceil N\text{Pr}_f \rceil$ times of overestimation. Hence, an upper bound on the probability of correct identification of each level is
\begin{equation}
\text{Pr}_u=\sum_{i=0}^{\lceil N\text{Pr}_f \rceil}{\binom{K}{i} \text{Pr}_{o}^{i}\left( 1-\text{Pr}_o \right) ^{K-i}} \label{eq35}
\end{equation}
where $\binom{K}{i} = \frac{K!}{i!\left( K-i \right) !}$ and $K$ represents the number of estimations at each level. From {\emph{Theorem 5}} in \cite{the_num_det_RMT}, 
\begin{equation}
\lim_{N_b\rightarrow \infty}\text{Pr}\left\{ \mathbf{\hat{p}}\left( k \right) =2m_k \right\} =1 \label{eq36}
\end{equation}
hence, $\lim_{N_b\rightarrow \infty}\text{Pr}_o=0$. Then, we have
\begin{equation}
\lim_{\text{Pr}_o\rightarrow 0}\text{Pr}_u= \lim_{\text{Pr}_o\rightarrow 0}\left( 1-\text{Pr}_o \right) ^K=1 . \label{eq37}
\end{equation}

\section{Simulation Results}

\subsection{Simulation Setup}
Monte Carlo simulation results are conducted to evaluate the performance of the proposed algorithm. Unless otherwise mentioned, we consider an SFBC-OFDM system with $N_r = 8$ receive antennas, $N=128$ sub-carriers, cyclic prefix length $\nu =10$, and 4-PSK modulation. The channel is assumed to be frequency-selective and consists of ${L_h} = 6$ statistically independent taps, each modeled as a zero-mean complex Gaussian random variable with an exponential power delay profile \cite{blind_SFBC}, $\sigma_t^2 = e^{-t/5}$. The probability of false alarm, ${\Pr _f}$ was set to ${10^{ - 4}}$ and the number of observed OFDM symbols ${N_b}$  was 100. The average probability of correct identification ${{\rm{Pr}}}$ was used as a performance measure, defined as
 \begin{equation}
 \text{Pr}=\frac{1}{7}\sum{\text{Pr}\left\{ C|C \right\}}. \label{eq38}
 \end{equation}
 The SFBC pool is set to $\{ \text{SA}$, $\text{SM}^{\left( 2 \right)}$, $\text{AL}$, $\text{SFBC}^{(1)}$, $\text{SFBC}^{(2)}$, $\text{SFBC}^{(3)}$, $\text{SM}^{\left( 3 \right)}\}$. Simulation of each code was run for 1000 trials.

\subsection{Performance Evaluation}

\subsubsection{Case of two transmit antennas}

\begin{figure}
  \centering
  \includegraphics[width=0.5\textwidth]{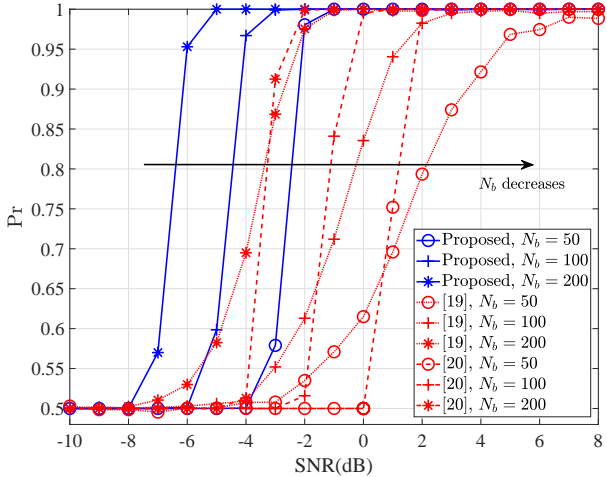}\\
  \caption{Performance comparison of the proposed algorithm and the algorithms in \cite{blind_SFBC,My_paper_WCNC} for $N_t=2$ and different $N_b$ on the average probability of correct identification $\rm Pr$, where the receiver has $N_r = 8$ antennas and the channel consists of $L_h = 6$ independent taps which follow a zero-mean complex Gaussian random variable with an exponential power delay profile, $\sigma_t^2 = e^{-t/5}$. }\label{fig4}
\end{figure}

Actually, \cite{blind_SFBC} only identified commonly used SFBCs, $\rm AL$ and ${\rm SM}^{(2)}$. The number of receive antennas in \cite{blind_SFBC} is also relatively small ($N_r=2$ to $5$). Indeed, that is an advantage for this algorithm. For a fair comparison in this paper, we assume that the number of transmit antennas is 2 and known by the receiver and the receiver has $N_r=8$ antennas. This is a reasonable assumption because in some situations, for example, military applications, additional antennas can be used to improve the performance. Fig. \ref{fig4} shows the performance of the proposed algorithm in comparison with the algorithms in \cite{blind_SFBC} and \cite{My_paper_WCNC}. The set of time lags $\varUpsilon$ in \cite{blind_SFBC} was set to $\{0, 1, 2, 3, 4, 5, 6 \}$ with cardinality $\left| \varUpsilon \right|=7$ in case the performance is restricted.  The results show that our proposed algorithm outperforms the algorithms in \cite{blind_SFBC} and \cite{My_paper_WCNC}. A 3-4 dB performance gain results from the proposed algorithm in comparison with \cite{My_paper_WCNC}, which reflects a more accurate estimation by utilizing the asymptotically accurate expression in \eqref{eq26} and the special distance formula in \eqref{eq33}. Fig. \ref{fig4} also shows that the proposed algorithm significantly outperforms the algorithm in \cite{blind_SFBC} for a very short observation period.

\subsubsection{Unknown number of transmit antennas and effect of the number of OFDM symbols}
\begin{figure}
  \centering
  \includegraphics[width=0.5\textwidth]{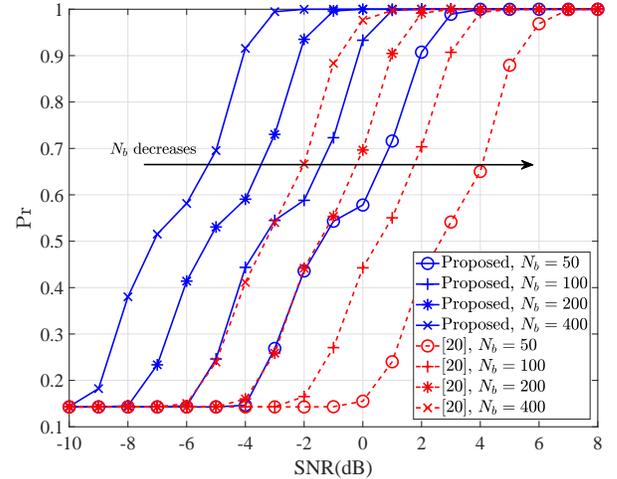}\\
  \caption{Effect of the number of OFDM symbols, ${N_b}$, on the average probability of correct identification $\rm Pr$. }\label{fig6}
\end{figure}

The algorithm in \cite{blind_SFBC} cannot support a large SFBC pool. Next, a comparison between the proposed algorithm and the algorithm in \cite{My_paper_WCNC} is provided. Fig. \ref{fig6} shows the compared performance and the average probability of correct identification of the proposed algorithm for ${N_b}=50$, ${N_b}=100$, ${N_b}=200$, ${N_b}=400$. As expected, the proposed algorithm outperforms the algorithm in \cite{My_paper_WCNC} significantly by about 2.5-3.5 dB. Additionally, the performance of the proposed algorithm improves with increasing ${N_b}$ since a more accurate Tracy-Widom distribution is achieved. It is noteworthy that the performance of the proposed algorithm for $N_b=100$ is between that of the algorithm in \cite{My_paper_WCNC} for $N_b= 200$ and $400$. This result indicates that the proposed algorithm performs well even for a short observation period.

\begin{table*}
\centering
\caption{FLOPS comparison between the proposed algorithm, the algorithms in \cite{blind_SFBC,My_paper_WCNC} for four groups of parameters: \protect \\
I. $N=128,N_r=4,N_{b}=100$; II. $N=64,N_r=8,N_{b}=100$; III. $N=128,N_r=8,N_{b}=50$; IV. $N=128,N_r=8,N_{b}=100$.}
\label{table2}
\begin{tabular}{@{}ccccccc@{}}
\toprule
Identification algorithm                  & Main computational cost                                           & Group I     & Group II   &  Group III            & Group IV \\ \midrule
Proposed algorithm                        & $48N{N_r^3} + 24N{N_b}{N_r^2} $                         & 5,308,416 & {\bf 11,403,264} &12,976,128 & {\bf 22,806,528}                     \\
The algorithm in [19]             & $8{N_b}{\left|\varXi \right|}\left( {N + \nu } \right)\left( {\left| \Upsilon  \right| + 1} \right)$               & {\bf{5,299,200}} &         13,260,800 & {\bf 12,364,800} & 24,729,600 \\
The algorithm in [20]         &  $64N{N_r^3} + 32N{N_b}{N_r^2} $                  &7,077,888 &  15,204,352     &17,301,504& 30,408,704                      \\ \bottomrule
\end{tabular}
\end{table*}

\subsubsection{Evaluation of computational complexity}
Based on the number of floating point operations (flops) definitions in \cite{matrix_computations}, the main computational complexities of the proposed algorithm and the algorithms in \cite{blind_SFBC} and \cite{My_paper_WCNC} are given by $48N{N_r^3} + 24N{N_b}{N_r^2} $, $8{N_b}{\left|\varXi \right|}\left( {N + \nu } \right)\left( {\left| \Upsilon  \right| + 1} \right)$ and $64N{N_r^3} + 32N{N_b}{N_r^2} $, respectively. Here, the number of flops for the eigenvalue decomposition is $64N_r^3$ using the QR algorithm, and $\varXi$ denotes the set of receive antenna pairs defined as $\varXi =\left\{ \left( v_1,v_2 \right) :v_1\ne v_2,\ \text{and\ }v_1<v_2 \le N_r \right\} $. In the previous case, i.e., $N_r=8$, $N=128$, $\nu = 10$, $N_b=100$, $\left|\varXi \right| =28$ and $\left| \varUpsilon \right| =7$ and the proposed algorithm requires 22,806,528 flops.  Employing the TMS320C6678 processor (a Digital Signal Processor produced by Texas Instruments) with 160 Giga-flops \cite{DSP1}, the proposed algorithm requires only about 130 $\mu$s, while in the LTE standard, 7.14 ms are spent transmitting 100 OFDM symbols with one block duration of 71.4 $\mu$s\cite{sesia2009lte}. The execution times of other algorithms are summarized in Table \ref{table2}. We can see that the proposed algorithm has comparable computational complexity to the algorithm in \cite{blind_SFBC}, although it achieves better performance as shown in Fig. \ref{fig4}.
The proposed algorithm is also suitable for parallel implementation owing to the independence of eigenvalue decompositions at different sub-carriers. By employing field programmable gate arrays or CUDA-enabled graphics processing units, the computational complexity decreases $N$ times.

\subsection{Effect of the False Alarm Probability}

\begin{figure}
  \centering
  \includegraphics[width=0.5\textwidth]{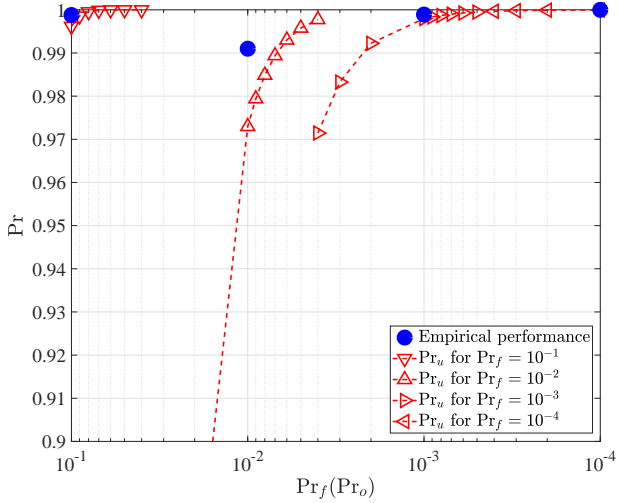}\\
  \caption{Effect of the false alarm probability, ${{\rm Pr}_f}$, on the average probability of correct identification $\rm Pr$.}\label{fig10}
\end{figure}

In Fig. \ref{fig10}, we use the circle marker to show the simulation performance of the average probability of correct identification for different false alarm probabilities (horizontal axis represents ${\rm Pr}_f$). The SNR is set to 6 dB. Four dotted lines represent joint ${\rm Pr}_u$ of the whole tree with the probability of overestimation ${\rm Pr}_o$ for ${\rm Pr}_f = 10^{-1}$, ${\rm Pr}_f = 10^{-2}$, ${\rm Pr}_f = 10^{-3}$ and ${\rm Pr}_f = 10^{-4}$, respectively, using Equation \eqref{eq35} (horizontal axis changes to ${\rm Pr}_o$). We can see that the empirical points are very close to and a little higher than the theoretical points of the upper bound since Equation \eqref{eq34} is an approximate formula which results in an error when substituting ${\rm Pr}_f$ into ${\rm Pr}_o$. The results indicate that the performance decreases when the ${\rm Pr}_f$ gets close to 0.01.
 
\subsection{Effect of the Number of OFDM Subcarriers}

\begin{figure}
  \centering
  \includegraphics[width=0.5\textwidth]{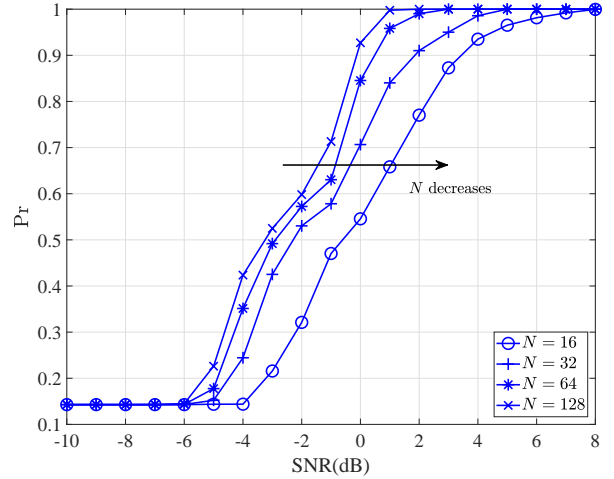}\\
  \caption{Effect of the number of OFDM sub-carriers, ${N}$, on the average probability of correct identification $\rm Pr$.}\label{fig7}
\end{figure}

Fig. \ref{fig7} presents the average probability of correct identification of the proposed algorithm for different numbers of sub-carriers, $N$. The performance of the proposed algorithm improves when increasing ${N}$ but with diminishing returns since the distance between the estimated ${\bf{\hat q}}$ and theoretical one converges rapidly with increasing $N$. In addition, a large $N$ results in a large computational complexity.

\subsection{Effect of the Number of Receive Antennas}

\begin{figure}
  \centering
  \includegraphics[width=0.5\textwidth]{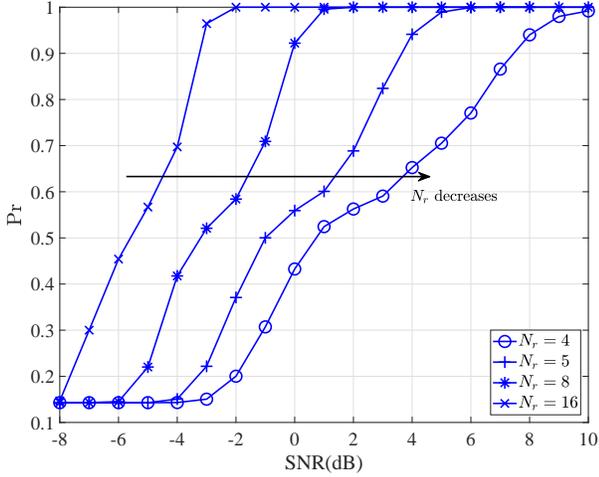}\\
  \caption{Effect of the number of receive antennas, ${N_r}$, on the average probability of correct identification $\rm Pr$.}\label{fig8}
\end{figure}

Fig. \ref{fig8} illustrates how the average probability of correct identification of the proposed algorithm is influenced by the number of receive antennas, ${N_r}$. With ${N_r}$ increasing, the performance of the proposed algorithm improves because the estimation of the noise variance in the denominator of Equation \eqref{eq25} and the expression in \eqref{eq26} become more accurate. 

\subsection{Effect of the Modulation Type}

\begin{figure}
  \centering
  \includegraphics[width=0.5\textwidth]{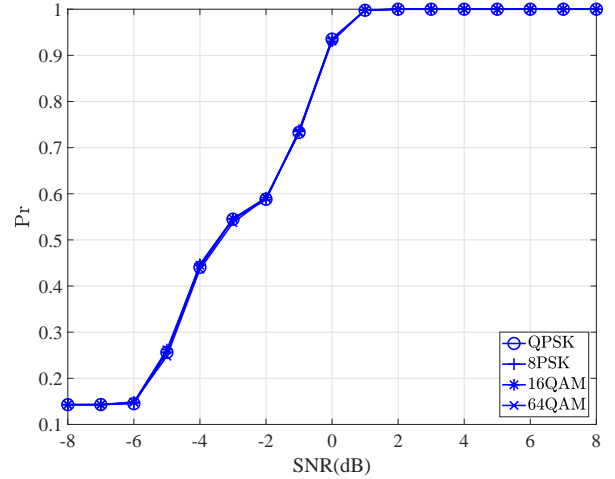}\\
  \caption{Effect of the modulation type on the average probability of correct identification $\rm Pr$.}\label{fig9}
\end{figure}

Fig. \ref{fig9} shows the average probability of correct identification of the proposed algorithm for different modulation types. The performance of the proposed algorithm does not depend on these modulation types, which are mandatory for most of the wireless standards. This is explained by the fact that the elements of the feature vector are determined by the matrix ${\bf M}_{k}$ in \eqref{eq22} and is independent of the modulation type. However, the proposed algorithm fails when the transmitter emits real modulation signals because we stack the real and imaginary part of the signals, and the imaginary part will be zero for real modulation.

\subsection{Effect of the Timing Offset}

Perfect timing synchronization was assumed in this paper. Here, we evaluate the performance of the proposed algorithm in the presence of a timing offset. The sample timing offset (STO) is modeled as in \cite{MIMO-OFDM_with_MATLAB}, which depends on the location of the estimated FFT window starting point of OFDM symbols. The effects of STO are classified into the following four different cases:
\begin{itemize}[leftmargin=*]
\item \emph{\textbf{Case I}}: The window starting point coincides with the exact timing;

\item \emph{\textbf{Case II}}: The window starting point is before the exact timing, yet after the end of the channel response to the previous OFDM symbol;

\item \emph{\textbf{Case III}}: The window starting point is estimated to exist prior to the end of the channel response to the previous OFDM symbol. In this case, the orthogonality among sub-carriers is destroyed by the inter-symbol interference (ISI);

\item \emph{\textbf{Case IV}}: The window starting point is after the exact point, hence, the received signal includes the ISI and inter-channel interference (ICI).

\end{itemize}

\begin{figure}
  \centering
  \includegraphics[width=0.5\textwidth]{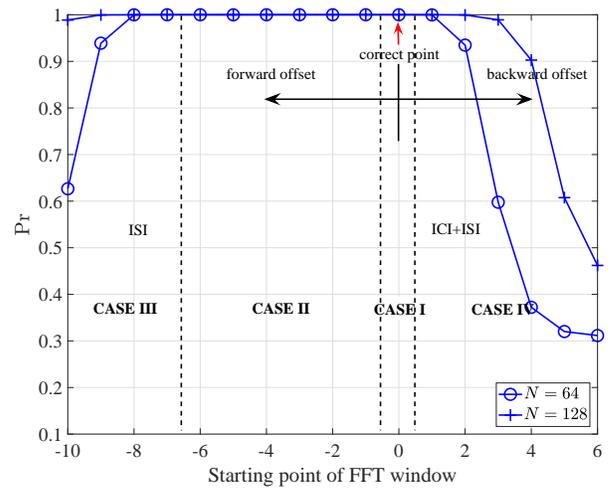}\\
  \caption{Effect of the starting point of FFT window on the average probability of correct identification $\rm P$ at SNR = 6 dB.}\label{fig11}
\end{figure}

Fig. \ref{fig11} illustrates the performance of the proposed algorithm at SNR = 6 dB for different STOs and values of $N$. One can notice that the proposed algorithm mostly identifies correctly for a small forward offset, as in \emph{\textbf{Case II}}, but fails for a large offset, as in \emph{\textbf{Case III}} and \emph{\textbf{Case IV}}, as the discriminating feature at an adjacent sub-carrier is destroyed by the ICI and ISI. However, the effect of the ISI will be dispersed under the condition of a large $N$ with the improvement of performance \cite{MIMO-OFDM_with_MATLAB}.

\subsection{Effect of the Frequency Offset}

\begin{figure}
  \centering
  \includegraphics[width=0.5\textwidth]{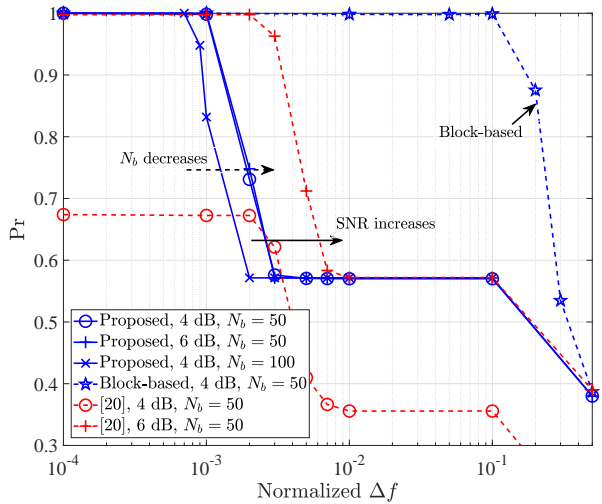}\\
  \caption{Effect of the normalized CFO on the probability of correct identification ${\rm Pr}$ for the proposed algorithm and the algorithms in \cite{blind_SFBC,My_paper_WCNC} at SNR = 0 dB and 6 dB.}\label{fig12}
\end{figure}

We consider the typical parameters of the LTE system to evaluate the impact of carrier frequency offset (CFO), with the number of sub-carriers being $N=128$ (the channel bandwidths are 1.4 MHz), and the number of processed OFDM symbols is equal to $N_b=50$ and $N_b=100$. The algorithm in \cite{My_paper_WCNC} are also compared with the proposed algorithm. The normalized carrier frequency offset is modeled as in \cite{MIMO-OFDM_with_MATLAB}. Fig. \ref{fig12} presents the effect of the normalized CFO to the sub-carrier spacing 15 kHz, $\Delta f$, on the performance at SNR = 4 dB and 6 dB. The results in Fig. \ref{fig12} show that the proposed algorithm is robust for $\Delta f < {10^{ - 3}}$ and outperforms the algorithm in \cite{My_paper_WCNC} at a relatively low SNR for $N_b = 50$. Additionally, the proposed algorithm performs better for a reduced observation period since the ICI destroys the orthogonality of SFBCs\cite{ICI_affects_OSTBC} and this impact accumulates with increasing the number of the processed OFDM symbols. From a practical point of view, with the successful estimation of the starting points of OFDM symbols, we can use a block-based process that operates on each OFDM symbol by removing the CP and doing the FFT operation, and then aligning the starting point of the next OFDM symbol to ease the impact of CFO for a blind receiver. Fig. \ref{fig12} shows the significant performance improvement using this block-based process. Furthermore, we can use a blind frequency offset compensation technique \cite{blind_freq_offset} by utilizing the kurtosis-type criterion before OFDM domodulation to reduce the effect of the frequency offset.

\subsection{Effect of the Doppler}

\begin{figure}
  \centering
  \includegraphics[width=0.5\textwidth]{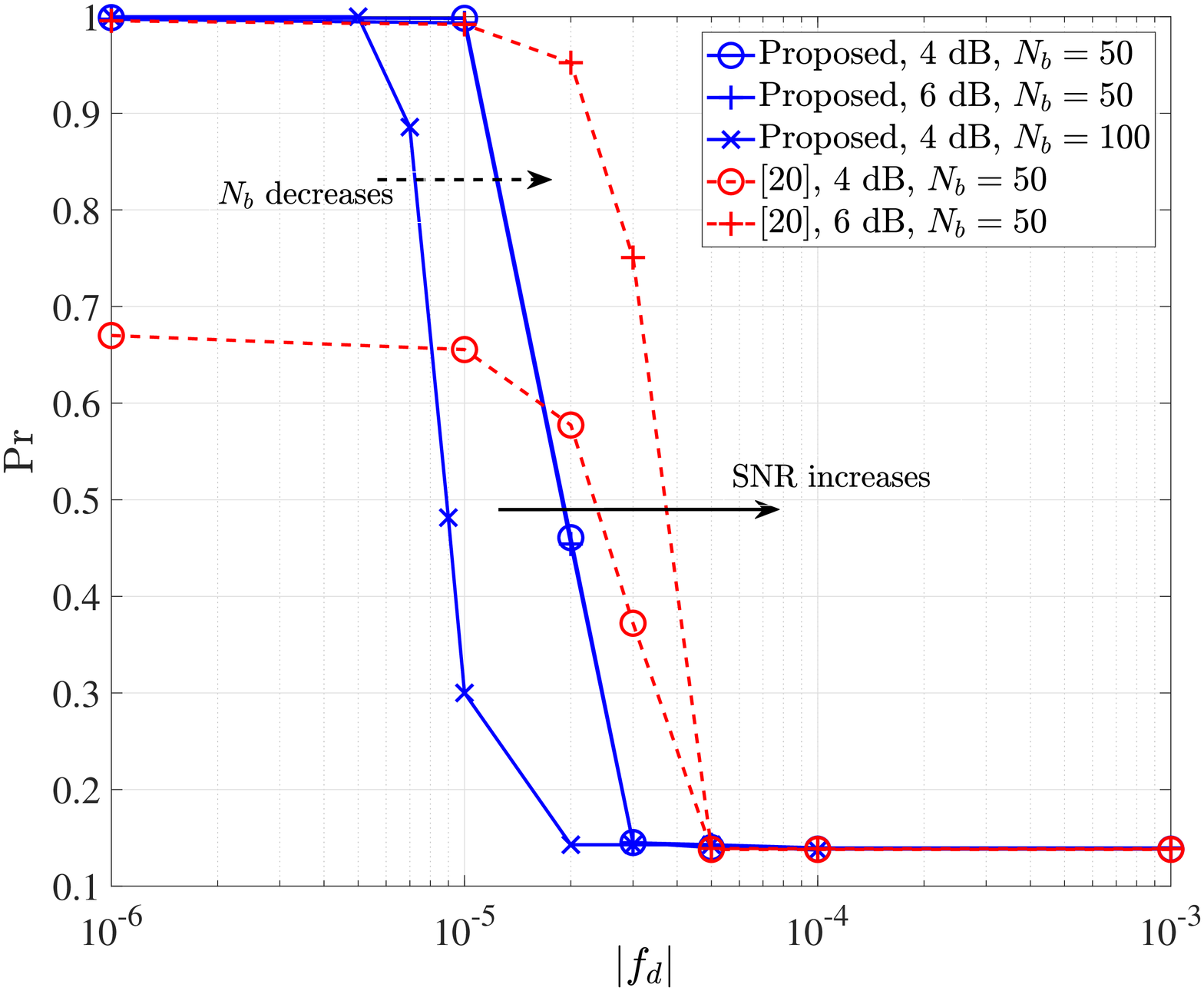}\\
  \caption{Effect of the maximum Doppler spread on the probability of correct identification ${\rm Pr}$ for the proposed algorithm and the algorithms in \cite{blind_SFBC,My_paper_WCNC} at SNR = 0 dB and 6 dB.}\label{fig13}
\end{figure}

The previous analysis assumed static channels over the observation period. Typical parameters of the LTE standard with the channel bandwidth of 1.4 MHz ($N=128$), sampling rate of 1.92 MHz and the number of processed OFDM symbols being $N_b = 50$ and $N_b=100$are assumed here to evaluate the impact of the Doppler frequency on the performance of the proposed algorithm and the algorithm in \cite{My_paper_WCNC}. Fig. \ref{fig13} shows the probability of correct identification versus the maximum Doppler frequency normalized to the sampling rate, $|f_d|$, at SNR = 4 dB and 6 dB. The results show that the proposed algorithm outperforms the algorithm in \cite{My_paper_WCNC} in the low-SNR regime with a small Doppler spread for a reduced observation period, and is robust for $ |f_d| < {10^{ - 5}}$ when $N_b=50$. 

\section{Conclusions}
We proposed a novel algorithm to identify SFBC-OFDM signals over frequency-selective channels. The dimension of the signal subspace of the received signals at adjacent sub-carriers is proposed to be the discriminating feature after the analysis of the received signal subspace. Then, we construct a feature vector to classify different SFBCs, whose elements are estimated by using a serial binary hypothesis test based on an asymptotically accurate RMT expression. Furthermore, a decision tree and a special distance metric are proposed to reduce the computational complexity and improve the performance, respectively. The proposed algorithm does not need prior information about the number of transmit antennas, channel coefficients, modulation mode and noise power. The simulations demonstrated that the enhanced identification performance and reduced computational complexity are achieved under frequency selective fading with a short observation period. Future works include devising robust identification of SFBC schemes to address the effect of the frequency offsets and Doppler spreads.

\section*{APPENDIX}

The orthogonal SFBC$^{(1)}$ of rate $\frac{1}{2}$ using ${N_t} = 3$ transmit antennas is defined by the following coding matrix\cite{STBC_Tarokh}
\begin{equation}
\mathbf{C}^{{\rm{SFBC}^{(1)}}}\left( \mathbf{x}_b \right) =\left[ \begin{matrix}
	x_{b,0}&		x_{b,1}&		x_{b,2}\\
	-x_{b,1}&		x_{b,0}&		-x_{b,3}\\
	-x_{b,2}&		x_{b,3}&		x_{b,0}\\
	-x_{b,3}&		-x_{b,2}&		x_{b,1}\\
	x_{b,0}^{*}&		x_{b,1}^{*}&		x_{b,2}^{*}\\
	-x_{b,1}^{*}&		x_{b,0}^{*}&		-x_{b,3}^{*}\\
	-x_{b,2}^{*}&		x_{b,3}^{*}&		x_{b,0}^{*}\\
	-x_{b,3}^{*}&		-x_{b,2}^{*}&		x_{b,1}^{*}\\
\end{matrix} \right] ^T.
\end{equation}

The orthogonal SFBC$^{(2)}$ of rate $\frac{3}{4}$ using ${N_t} = 3$ transmit antennas is defined by the following coding matrix\cite{Larsson_STBC}
\begin{equation}
{{\bf{C}}^{{\rm SFBC}^{(2)}}}\left( {{{\bf{x}}_b}} \right) = \left[ {\begin{array}{*{20}{c}}
{{x_{b,0}}}&0&{{x_{b,1}}}\\
0&{{x_{b,0}}}&{x_{b,2}^*}\\
{ - x_{b,1}^*}&{ - {x_{b,2}}}&{x_{b,0}^*}
\end{array}\quad \begin{array}{*{20}{c}}
{ - {x_{b,2}}}\\
{x_{b,1}^*}\\
0
\end{array}} \right].
\end{equation}

Last, the orthogonal SFBC$^{(3)}$ of rate $\frac{3}{4}$ using ${N_t} = 3$ transmit antennas is defined by the following coding matrix\cite{Larsson_STBC}
\begin{equation}
{{\bf{C}}^{{\rm SFBC}^{(3)}}}\left( {{{\bf{x}}_b}} \right) = \left[ {\begin{array}{*{20}{c}}
{{x_{b,0}}}&{ - x_{b,1}^*}&{x_{b,2}^*}\\
{{x_{b,1}}}&{x_{b,0}^*}&0\\
{{x_{b,2}}}&0&{ - x_{b,0}^*}
\end{array}\quad \begin{array}{*{20}{c}}
0\\
{ - x_{b,2}^*}\\
{x_{b,1}^*}
\end{array}} \right].
\end{equation}

\ifCLASSOPTIONcaptionsoff
  \newpage
\fi

\bibliographystyle{IEEEtran}
\balance
\bibliography{./gao}

\end{document}